# Negative Thermal Expansion Behaviour in MZrF$_6$ (M=Ca, Mg, Sr): Ab-initio Lattice Dynamical Studies

M. K. Gupta[1], Baltej Singh[1,2], R. Mittal[1,2] and S. L. Chaplot[1,2]

[1]*Solid State Physics Division, Bhabha Atomic Research Centre, Mumbai 400085*
[2]*Homi Bhabha National Institute, Anushaktinagar, Mumbai-400094*

The thermal expansion behavior of metal fluorides can be tuned by choosing appropriate metal cation. We present ab-initio lattice dynamical studies on the metal fluorides (CaZrF$_6$, MgZrF$_6$ and SrZrF$_6$) and identify the anharmonic phonon modes responsible for the negative thermal expansion in these materials. These phonons involve ZrF$_6$ polyhedral rotational motion, which leads to large transverse amplitude of the vibrations of the fluorine atom in the Zr-F-Zr bond. The compounds with larger anisotropy in the thermal amplitude of the fluorine atoms show larger NTE behaviour. This has enabled to understand the large variation in thermal expansion behaviour of these compounds at high temperature. The calculations also predict decrease of the frequency of these anharmonic phonons with increasing temperature.

## I. Introduction

Expansion of lattice on increase of temperature is a common physical phenomenon. However, some of the exotic solid materials contract upon heating and are referred as negative thermal expansion materials (NTE), e.g., water, invar alloy, silicon and germanium. The contraction may be isotropic or anisotropic depending upon the structure and bonding. The compounds ZrW$_2$O$_8$[1-4], ScF$_3$[5-10] and Cd(CN)$_2$[11] are the few famous materials which show isotropic negative thermal expansion behaviour.Large number of compounds like LiAlSiO$_4$[12], Ag$_3$Co(CN)$_6$[13], KMn[Ag(CN)$_2$]$_3$[14], ZnAu$_2$(CN)$_4$[15, 16], MCN (M=Ag, Au, Cu)[17] show anisotropic thermal expansion, where one of the direction shows negative thermal expansion behaviour. These compounds consist of polyhedral units and NTE arise due tocooperative rotation and deformation of these polyhedral units which contract open spaces in the lattice[18].

For application of a NTE material, it has to be stable under conditions of use and amenable to the fabrication of composite or bulk component. The main drawback of metal organic framework(MOF) including cyanide bridges framework NTE materials is that they are not easy to process into ceramic bodies. They are largely hygroscopic in nature which reduced the coefficient of NTE, hence are not suitable for ambient environment applications. The discovery of large NTE in fluorides based compounds open the way to overcome this problem in certain extent. In the fluorides based compounds the terminal entity is fluorine, in contrast to oxygen in metal oxide framework. The large difference between electronegativity between M--F in comparison to M--O leads to largely ionic character of bonding in these compounds and enhances the flexibility[19]. The difference in the nature of bonding also significantly changes the thermos physical properties in these compounds. A large NTE behaviour has been reported in $ScF_3$ compound. The compound is also transparent in infrared regime hence suitable for optical fibre application. At low temperature it transforms from cubic(Fd-3m) to rhombohedral (R3c) phase, this transformation is mediated by freezing of a R- point mode. Interestingly the same mode largely contributes in the NTE behaviour. Hence structural phase transition kills NTE behaviour in this compound. Most of the fluorites with $ReO_3$ structure show NTE behaviour in high temperature cubic phase, while in the low temperature phase NTE is reduced or absent. This might be due to phonon softening along R-M line in the Brillouin zone which is observed in the cubic phase.

Recently a large number of fluoride compounds $M^{II}Zr^{IV}F_6$ (M=Ca, Mg, Sr), which resemble $ReO_3$ structure are found to show large diversity in thermal expansion behaviour, i.e. from large negative expansion to large positive expansion. The studies show that in $MZrF_6$ series the thermal expansion can be tuned by choosing appropriate metal cation. $MgZrF_6$ has almost zero thermal expansion. Based on temperature dependent synchrotron X-ray total scattering diffraction the authors[20]proposed that flexibility of M···F atomic linkages in $MZrF_6$ governs the thermal expansions in these fluorites. The more flexible linkage favors the negative expansion, while stiffer linkage leads to positive expansion behaviour. Similar prediction has been made[19] in $M_2O$ oxides based on first-principles density functional theory calculations. The first observation of NTE behaviour in this series is made by Hancock et al[21]on CaZrF6 and $CaHfF_6$. The authors found that $CaZrF_6$ and $CaHfF_6$have much stronger negative thermal expansion (NTE) than in $ZrW_2O_8$ and other corner-shared framework structures. The NTE behavior is comparable with metal cyanides and metal−organic framework compounds. These ceramic compounds are transparent over a wide wavelength range and can be handled in air, which is not the case for cyanides. The coefficient of NTE in $CaZrF_6$ is strongly temperature dependent. The compound is relatively soft with bulk modulus of 37 GPa, and shows disorder at pressure about 400MPa. The NTE coefficient in $CaZrF_6$ is greater in magnitude than $ScF_3$ and comparable to metal cyanides[11, 13, 22] and $CarF_6$ is not hygroscopic.

The large diversity in thermal expansion behaviour observed by Lei Hu et al[20] among these compounds provides an opportunity to analyse the thermal expansion behaviour in terms of the bonding and local distortion as degree of freedom. However, there is lack of microscopic understanding of the diversity of thermal expansion behaviour in this series of compounds. In view of this we have performed first principles phonon study of variety of fluorites and analysed the factors which control the thermal expansion coefficient of these compounds. In the present paper, our focus is to understand the diversity in thermal expansion behaviour in these metal fluorides in terms of atomic dynamics. Here we present the lattice dynamical studies on the metal fluorides ($CaZrF_6$, $MgZrF_6$ and $SrZrF_6$)which show large variation in thermal expansion behaviour.

**II. COMPUTATIONAL DETAILS**

The VASP[23, 24] along with PHONON software[25] is used to calculate the phonon frequencies in entire Brillouin zone. The software requires the forces on various atoms after displacement along various directions. Which are obtained by computing the Hellman-Feynman forces on various atom in various configuration on a supercell with various (±x, ±y, ±z) atomic displacement patterns. The Hellman-Feynman forces on the atoms in the super cell have been calculated using density functional theory as implemented in the VASP software[23, 24]. The $A^{II}B^{IV}F_6$ in cubic phase(space group 225 Fd-3m) has 3 symmetrically inequivalent atom in unit cell(space group 225 Fm-3m), and total number of atom in primitive unit cell is 8. The force calculations is performed by displacing the symmetrically inequivalent atoms along (±x,±y,±z) direction. The energy cutoff is 580 eV and a 8 ×8 × 8 k point mess has been used to obtain energy convergence in total energy of the order of meV, which is sufficient to obtain the required accuracy in phonon energies. The Monkhorst Pack method is used for k point generation[26]. The exchange-correlation contributions were included within PBE generalized gradient approximation (GGA)[27]. The convergence criteria for the total energy and ionic forceswere set to $10^{-8}$ eV and $10^{-3}$ eV Å$^{-1}$, respectively.

In the quasi-harmonic approximation, the volume thermal expansion coefficient [] of a crystalline material, is given by the following relation: $\alpha_V = \frac{1}{BV}\sum_i \Gamma_i C_{Vi}(T)$. Here $\Gamma_i = -\frac{V}{E_i}\frac{dE_i}{dV}$ is the mode Grüneisen parameter, which is a measure of the volume dependence of the phonon frequency. $C_{Vi}(T)$ is the specific heat contribution of the $i^{th}$ phonon mode (of energy $E_i$) at temperature T, while B and V stand for the bulk modulus and volume of the compound, respectively. In the above relation, all the quantities except $\Gamma_i$are positive at all temperatures. Therefore, the sign and magnitude of $\Gamma_i$ govern the thermal expansion of the lattice, while the phonon energy range over which $\Gamma_i$ is negative determines the temperature range over which the material will show NTE.

## III. RESULTS AND DISCUSSION

### A. Phonon Dispersion Relation and Phonon Density of States

We have calculated (Fig1) the phonon dispersion relation along various high symmetry directions for $CaZrF_6$, $MgZrF_6$ and $SrZrF_6$. The calculated phonon dispersion relation shows that low energy phonons below 30 meV in $MgZrF_6$ shows large dispersion, while the dispersion is least in $SrZrF6$. The high energy dispersion in all the three compounds lies between 60 to 80 meV. The longitudinal and transverse acoustic branches along Γ-X line show large slope in $MgZrF_6$ and least in $SrZrF_6$. This indicates that $MgZrF_6$ is the hardest material among the three fluorides. The four dispersion branches in $SrZrF_6$ form a narrow band around 25-30meV, however in $CaZrF_6$ and $MgZrF_6$ compounds these dispersion branches show large dispersion in a wide energy range from 25 to 50meV.

In Fig 2 We have shown the calculated atomic and total density of states. The first peak in the partial density of states of fluorine atom in $MgZrF_6$ is at about 12meV, however, in $CaZrF_6$ and $SrZrF_6$ the peak is at about 8 meV. The shift in the fluorine density of states peak in $MgZrF_6$ is attributed to difference in nature of bonding of fluorine atom with Zr in these compounds. The metal atom (M=Ca, Mg, Sr) contribution in density of states is extended up to 40meV, 50 and 30 meV for Ca, Mg and Sr compound respectively. Interestingly in $SrZrF_6$ the Sr density of states shows huge localized density of states at about 18 meV. The zirconium atom shows contribution in the entire spectrum of density of states in all the three compounds. The lowest peak in the partial density of state due to Zr is at about 10 meV in both $CaZrF_6$ and $SrZrF_6$, however it is at 12 meV in $MgZrF_6$. This may be again due to difference in the strength of Zr-F bond among these compounds. The total density of states is weighted sum of atomic density of states which also show that the lowest peak in the density of state is at about 8 meV in both $CaZrF_6$ and $SrZrF_6$ and 12 meV in $MgZrF_6$. The total density of state shows band gap at 35-57meV, 50-57meV and 40-57meV in the Ca, Mg and Sr compounds respectively. The individual contribution of atoms is important to understand the NTE behaviour, as we will see in next section that NTE in these fluorides are dominated by the fluorine atom dynamics.

## B. Thermal Expansion Behaviour

The thermal expansion behaviour in all the three fluorides has been computed under the quasiharmonic approximation. The pressure dependece of phonon frequencies has been used to calculate the mode Gruneisen parameters, which arefurther used to compute the thermal expansion coefficient.

The calculated Gruneisen parameters of the three fluorides are shown in Fig 3. The Gruneisen parameter is calculated as a function of phonon energy averaged over entire Brillouin zone. The modes below 10 meV shows negative Gruneisen parameters. For $CaZrF_6$, the maximum negative value of the Gruneisen parameters is -26 forphonons of 5 meV energy. However for $MgZrF_6$ and $SrZrF_6$the maximum negative value of -18 and -12 respectively at about 5 meV. It seems that the phonon modes of energy around 5meV have a major role to contract the lattice in all the three compouds. However, the value of coefficient of NTE would dependes on population of ~5meV phonons in these compounds.

The calculated Gruneisen parameters and specific heat contribution from each mode are used to compute the volume thermal expansion coefficient (Fig 4). As discussed above, the large negative Gruneisen parameters of low energy modes around 5 meV result in NTE behaviourin all the three compounds. The compounds $CaZrF_6$ shows largest negative volume thermal expansion coefficient at ~ 100K and NTE remainsup to a very large range of tempertaure. $SrZrF_6$ also shows similar behaviour but the magnitude of NTE coefficient is less in comparison to that in $CaZrF_6$. However, $MgZrF_6$ shows NTE behaviour only upto 300 K, and at high tempertures the coefficient of thermal expansion has a very small positive value (~$10^{-7}$ $K^{-1}$). Thedifference in behaviour as observed in $MgZrF_6$ in comparison to that of$CaZrF_6$ and $MgZrF_6$may be due to significant difference in the phonon spectra (Figs. 1 and 2). To understand this we have calculated (Fig 5)the contribution of phonon energy E averaged over Brillouin zone to the volume thermal expansion at 300K.

As mentioned in the computational section, the value of volume thermal expansion coefficient depends on sum of the product of specific heat and Gruneisen parameters. It seems that larger the negative value of Gruneisen parameters larger would be the negative coefficient of thermal expansion.For $MgZrF_6$ and $SrZrF_6$the maximum negative value of $\Gamma$ of phonons at about 5 meV is -18 and -12 respectively. However, this gives maximum negative thermal expansion coefficient of about -18×$10^{-6}$ and -42×$10^{-6}$ for Mg and Sr compoundsrespectivlelyat 100 K. This is not an anomaly but this is due to the difference in phonon density of states at 5meV. At 5meV, the phonon spectra of $MgZrF_6$ (Fig. 2) has a lower phonon denisty of states in comparison to that of SrZrF$_6$. Hence the specific heat contribution from 5meV phonons would be less in MgZrF$_6$.

We have compared the calculated fractional volume change with available experimental values in Fig 6. The thermal expansion behaviour of MgZrF$_6$ shows very good agreement with the availbale experimental data[28]. The experminetal data of SrZrF$_6$ are not availbale. However, in case of CaZrF$_6$ it shows good agreement with calculation only upto 200 K and after that it deviates significantly. This descrepancy might be attributed to the temperature dependent anharmonicity of phonons which is not taken care by quasiharmonic approximation. In order to understand this we have calculated the temperature dependence of phonon frequency of a few phonon modes in this compounds.

Thermal expansion in these compounds is argued to be related with large thermal amplitude of linkage atom between the polyhedras (Zr-F-Zr), which lies along (100) direction of the cubic cell. Hence we have calculated the anisotropic mean square displacements of various atoms contributed from phonons in entire Brillouin zone in these compounds (Fig 7(a) and (b)).The terminal atom i.e. fluorine has the maximum mean square displacement. As expected due to heavy mass the Zr atom has least values of mean squared displacement. Among the three compounds, the fluorine atom in SrZrF$_6$ compound shows the largest mean square displacement, which suggest that the nature of bonding in SrZrF$_6$ is softer in comparison toother two compouds, as discussed in the next section. Further the mean square amplitude is more along the perpendicular to the Zr-F-Zr bond in all three compounds (Fig7(b)). Although, the largest value of the mean square displacement in SrZrF$_6$, CaZrF$_6$ shows the maximum NTE behaviour. It should be noted that the ratio of anisotropic mean square displacement ($u_\perp^2/u_\parallel^2$) is 4.7, 4.4 and 4.2 at 200K in Ca, Sr and Mg compounds respectively (Fig7(b)). The large amplitude of the vibrations perpendicular to the Zr-F bond may correspond to librational motion of ZrF6 polyhedra.The anisotropy could drive the roational dynamics of ZrF$_6$polyhedra, which in turn could result in NTE behaviour in these compounds. This may explain the trends of thermal expansion behaviour in these fluorides.

The displacement pattern of Γ and X point phonon modes are shown in Fig 8. Both the modes show anti-phase rotation of ZrF$_6$ and Ca/Mg/SrF$_6$ units The Grüneisen parameter of the Γ-point optic mode is -50, -45 and -36 and for X point mode the values are -61, -47 and -35 for Ca, Mg and Sr compounds respectively. It seems that the rotational dynamics of the polyhedral leads to the maximum NTE in CaZrF$_6$ which is consistent with the observation; and it gives the least NTE in SrZrF$_6$ compound. However, SrZrF$_6$ has larger NTE than MgZrF$_6$, so in SrZrF6 something more than the rotational dynamics is adding to the NTE. This can be understood by calculating the polyhedral volume of various units in these compounds. The volume of the various polyhedral unit in these compounds are 15.6, 10.7, 19.3 and 11.2 Å$^3$ for CaF$_6$, MgF$_6$, SrF$_6$ and ZrF$_6$ respectively. In case MgZrF$_6$, both the polyhedral units have similar volume. However, in case of SrZrF$_6$/CaZrF$_6$, the SrF$_6$/CaF$_6$ polyhedral volume (19.3/15.6Å$^3$) is much larger than that of ZrF$_6$ (11.2 Å$^3$). The large difference is volume of the polyhedral units in Ca and Sr compounds may be responsible for polyhedral distortion which would ease in reducing the volume.

**C. Anharmonicity and Temperature Dependence of Phonon Modes**

The phonon frequencies are usually expected to soften with increase of temperature. However, in a few systems some of the phonons harden[2]. The shift of phonon frequency is due to phonon-phonon interaction and also due to anharmonic vibrations of atoms corresponding to a given phonon. It is difficult to estimate the shift of phonon frequency due to phonon-phonon interaction in entire Brillouin zone; however, one can make certain assumptions and can estimate the shifts of phonon energy as a function of temperature. We have calculated the potential energy profile of a few selected high symmetry point phonons in the Brillouin zone and used it to estimate the temperature dependence of the phonon frequencies. The procedure to calculate the temperature dependence of the phonon modes is given in Refs.[29, 30].

The lowest optical branch in all the three halides contributes to the negative thermal expansion behaviour. Interestingly this branch along Γ-X direction is dispersion less and provides large contribution in the density of states. Hence it significantly contributes in the thermodynamic behavior of MZrF$_6$ series. The lowest optic mode at the zone centre involves (Fig. 8) out of phase rotational motion of ZrF$_6$ and CaF$_6$ units, which lead to contraction in the cell dimension by utilizing the open space among these polyhedral units. However, along X-L and L-W this branch does not show significant softening. In order to see the effect of vibrational amplitude of such phonon modes, we have calculated the potential energy profile of these phonon modes. The crystal potential energy is then fitted to the expression $V(\theta_j) = a_{0,j} + a_{2,j}\theta_j^2 + a_{3,j}\theta_j^3 + a_{4,j}\theta_j^4$ (where θ$_j$ is the normal coordinate of the j$^{th}$ phonon mode and a$_{2,j}$, a$_{3,j}$, and a$_{4,j}$ are the coefficients of the harmonic and third and fourth order anharmonic terms, respectively). In Table I we have shown the calculated anharmonic coefficient (quartic anharmonicity) of a few selected phonons in all the three halides. We observe that the lowest both the optic mode at Γ-point and X point in the Brillouin zone show large quartic anharmonicity (Fig 9).

The temperature dependence of these anharmonic modes arise from two factors, namely, (1) the implicit anharmonicity due to volume change, and (2) the explicit anharmonicity due to increase in vibrational amplitudes. We find that in case of $CaZrF_6$ and $SrZrF_6$ for both the modes, the shift in phonon frequency due to implicit anharmonicity dominates over explicit anharmonicity, and the modes shows decrease in energy with increase of temperature. In case of $MgZrF_6$, the change in volume with temperature (Fig 5) is very small compared to $CaZrF_6$ and $SrZrF_6$ compounds; hence, the change in phonon frequency due to volume change at high temperature is small and explicit anharmonicity dominates.

## IV. Conclusions

We have extensively investigated the phonon dynamics in three isostructural $ReO_3$-type compounds. The large volume anharmonicity of low energy modes in $CaZrF_6$ and $SrZrF_6$ is responsible for NTE in these compounds. In $MgZrF_6$, at high temperature, the thermal expansion coefficient become much smaller due to some of the high energy phonon modes which contribute to the positive expansion behaviour and kill the NTE. The anisotropy in the mean square displacement ($u_\perp^2/u_\parallel^2$) correlates well with the thermal expansion behaviour in these compounds. Larger the ratio, higher is the NTE coefficient. We also predict the phonon frequency shifts at high temperature, where the softening of the anharmonic phonon modes due to large implict volume anharmonicty is partly compensated by explct anharmonicity due to increase of vibrational amplitude at high temperature.

**Acknowledgements**

S. L. Chaplot would like to thank the Department of Atomic Energy, India for the award of Raja Ramanna Fellowship. The use of ANUPAM super-computing facility at BARC is acknowledged.

[1] M. K. Gupta, R. Mittal, and S. L. Chaplot, Physical Review B **88**, 014303 (2013).

[2] G. Ernst, C. Broholm, G. Kowach, and A. Ramirez, Nature **396**, 147 (1998).

[3] C. Perottoni and J. Da Jornada, Science **280**, 886 (1998).

[4] T. Mary, J. Evans, T. Vogt, and A. Sleight, Science **272**, 90 (1996).

[5] A. van Roekeghem, J. Carrete, and N. Mingo, Physical Review B **94**, 020303 (2016).

[6] L. Hu, J. Chen, A. Sanson, H. Wu, C. Guglieri Rodriguez, L. Olivi, Y. Ren, L. Fan, J. Deng, and X. Xing, Journal of the American Chemical Society **138**, 8320 (2016).

[7] P. Lazar, T. Bučko, and J. Hafner, Physical Review B **92**, 224302 (2015).

[8] C. W. Li, X. Tang, J. A. Munoz, J. B. Keith, S. J. Tracy, D. L. Abernathy, and B. Fultz, Physical review letters **107**, 195504 (2011).

[9] B. K. Greve, K. L. Martin, P. L. Lee, P. J. Chupas, K. W. Chapman, and A. P. Wilkinson, Journal of the American Chemical Society **132**, 15496 (2010).

[10] S. U. Handunkanda, E. B. Curry, V. Voronov, A. H. Said, G. G. Guzmán-Verri, R. T. Brierley, P. B. Littlewood, and J. N. Hancock, Physical Review B **92**, 134101 (2015).

[11] V. E. Fairbank, A. L. Thompson, R. I. Cooper, and A. L. Goodwin, Physical Review B **86**, 104113 (2012).

[12] B. Singh, M. K. Gupta, R. Mittal, M. Zbiri, S. Rols, S. J. Patwe, S. N. Achary, H. Schober, A. K. Tyagi, and S. L. Chaplot, Journal of Applied Physics **121**, 085106 (2017).

[13] A. L. Goodwin, M. Calleja, M. J. Conterio, M. T. Dove, J. S. Evans, D. A. Keen, L. Peters, and M. G. Tucker, Science **319**, 794 (2008).

[14] K. Kamali, C. Ravi, T. Ravindran, R. Sarguna, T. Sairam, and G. Kaur, The Journal of Physical Chemistry C **117**, 25704 (2013).

[15] A. Goodwin, J. Am. Chem. Soc. **131**, 6334 (2009).

[16] M. K. Gupta, B. Singh, R. Mittal, M. Zbiri, A. B. Cairns, A. L. Goodwin, H. Schober, and S. L. Chaplot, Physical Review B **96**, 214303 (2017).

[17] M. K. Gupta, B. Singh, R. Mittal, S. Rols, and S. L. Chaplot, Physical Review B **93**, 134307 (2016).

[18] R. Mittal, M. K. Gupta, and S. L. Chaplot, Progress in Materials Science **92**, 360 (2018).

[19] M. K. Gupta, R. Mittal, S. L. Chaplot, and S. Rols, Journal of Applied Physics **115**, 093507 (2014).

[20] L. Hu, J. Chen, J. Xu, N. Wang, F. Han, Y. Ren, Z. Pan, Y. Rong, R. Huang, and J. Deng, Journal of the American Chemical Society **138**, 14530 (2016).

[21] J. C. Hancock, K. W. Chapman, G. J. Halder, C. R. Morelock, B. S. Kaplan, L. C. Gallington, A. Bongiorno, C. Han, S. Zhou, and A. P. Wilkinson, Chemistry of Materials **27**, 3912 (2015).

[22] A. B. Cairns, J. Catafesta, C. Levelut, J. Rouquette, A. Van Der Lee, L. Peters, A. L. Thompson, V. Dmitriev, J. Haines, and A. L. Goodwin, Nature materials **12**, 212 (2013).

[23] G. Kresse and D. Joubert, Physical Review B **59**, 1758 (1999).

[24] G. Kresse and J. Furthmüller, Physical Review B **54**, 11169 (1996).

[25] K. Parlinksi, 2003).

[26] H. J. Monkhorst and J. D. Pack, Phys. Rev. B **13**, 5188 (1976).

[27] J. P. Perdew, K. Burke, and M. Ernzerhof, Phys. Rev. Lett. **77**, 3865 (1996).

[28] J. Xu, L. Hu, Y. Song, F. Han, Y. Qiao, J. Deng, J. Chen, and X. Xing, Journal of the American Ceramic Society  (2017).

[29] B. Kuchta and T. Luty, The Journal of Chemical Physics **78**, 1447 (1983).

[30] N. Choudhury, S. L. Chaplot, and K. R. Rao, Physical Review B **33**, 8607 (1986).

TABLE I The calculated anharmonicity parameters in the potential of certain phonon modes. The value of the parameter '$a_{4,j}$' is extracted from fitting of $V(\theta_j) = a_{0,j} + a_{2,j}\theta_j^2 + a_{3,j}\theta_j^3 + a_{4,j}\theta_j^4$ to the potential energy profile of the mode with fixed value of '$a_{2,j}$' as determined from the harmonic phonon energies. '$a_{3,j}$' is found to be zero. The calculated Grüneisen parameter $\Gamma(E)$ of the respective mode is also given.

| Wave vector | E (meV) | $\Gamma(E)$ | $a_{2,j}$ ($10^{-3}$ eV) | $a_{4,j}$ ($10^{-6}$ eV) |
|---|---|---|---|---|
| CaZrF$_6$ | | | | |
| $\Gamma$ | 3.60 | -50 | 1.55 | 15.5 |
| X | 3.66 | -61 | 1.60 | 19.3 |
| MgZrF$_6$ | | | | |
| $\Gamma$ | 3.87 | -45 | 1.79 | 17.6 |
| X | 3.95 | -47 | 1.87 | 20.1 |
| SrZrF$_6$ | | | | |
| $\Gamma$ | 3.31 | -36 | 1.31 | 17.4 |
| X | 3.51 | -35 | 1.47 | 4.2 |

FIG. 1 (Color online) Calculated phonon dispersion relation of MZrF$_6$(M=Ca, Mg and Sr). The high symmetry points in the cubic Brillouin zone are defined as Γ=(0 0 0), X(1 0 0), L(1/2 1/2 1/2), W(1/2 1 0).

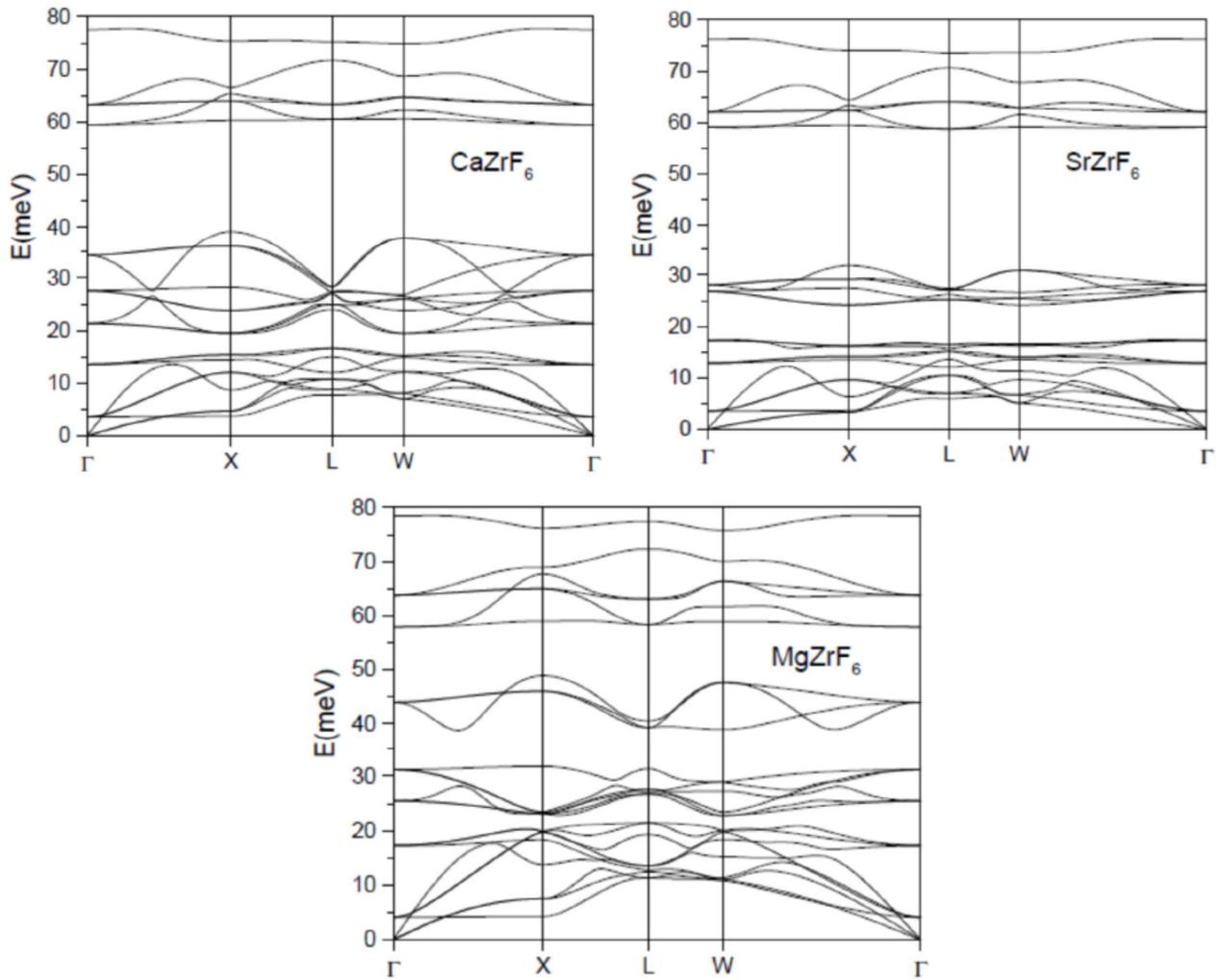

FIG. 2 (Color online) Calculated atomic partial phonon density of states of various atoms and the total phonon density of states in MZrF$_6$ (M=Ca, Mg and Sr).

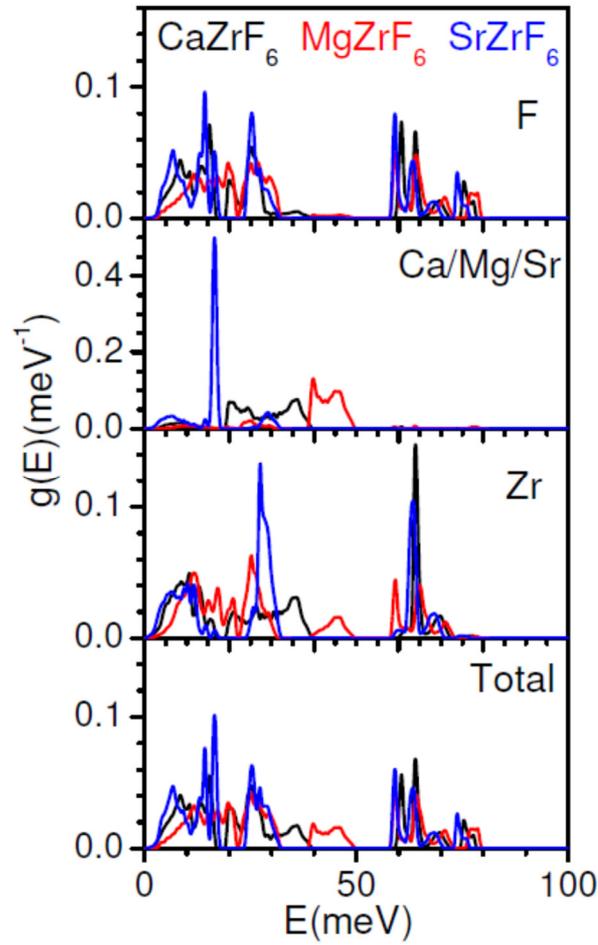

FIG. 3 (Color online) Calculated Grüneisen parameter as a function of phonon energy (E) of MZrF$_6$ (M=Ca, Mg and Sr); (a) plotted for individual phonons, and (b) averaged for all phonons around energy E.

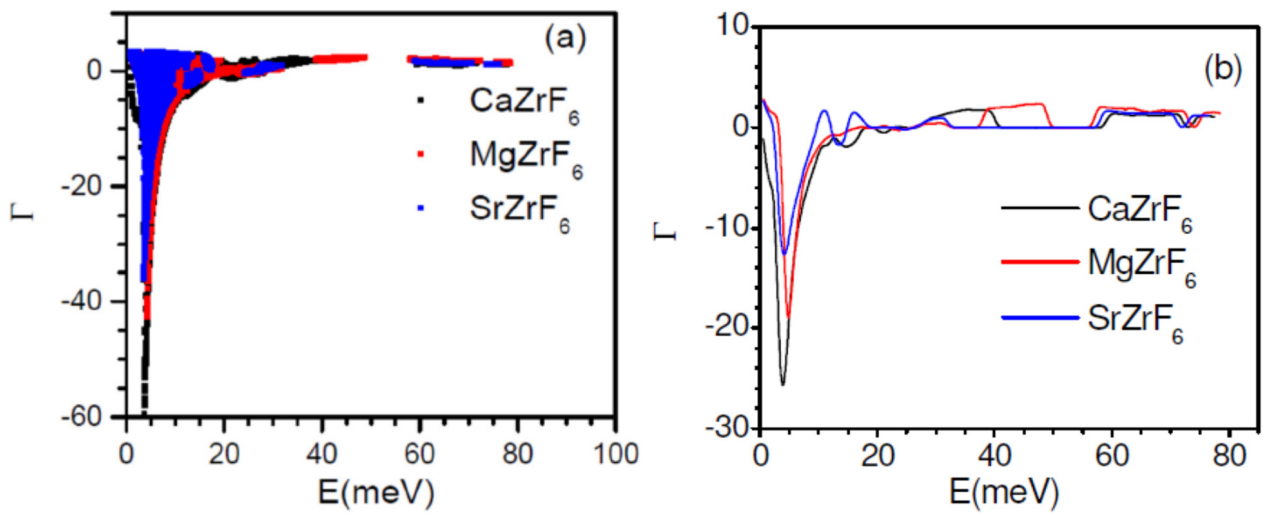

FIG. 4 (Color online) The calculated volume thermal expansion coefficient of $MZrF_6$(M=Ca, Mg and Sr) as a function of temperature.

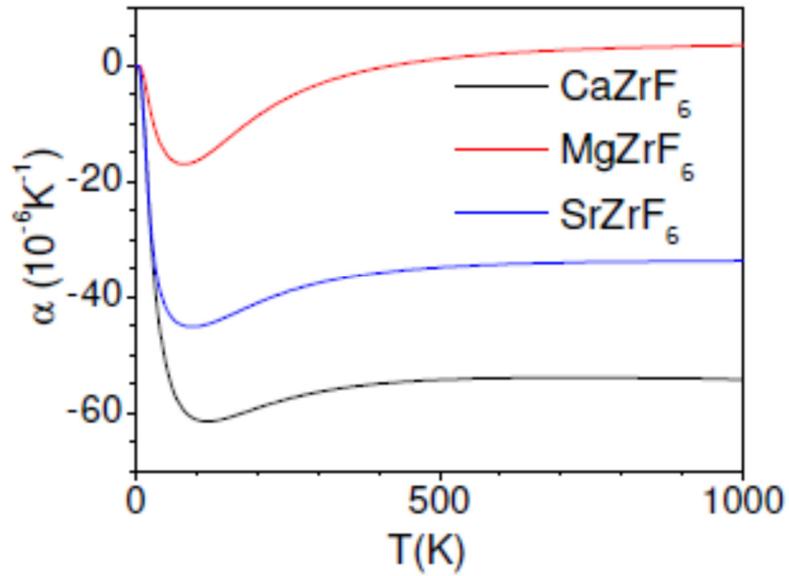

FIG. 5 (Color online) The contribution to the calculated volume thermal expansion coefficient of $MZrF_6$ (M=Ca, Mg and Sr) from phonons of energy E averaged over the Brillouin zone at 300 K.

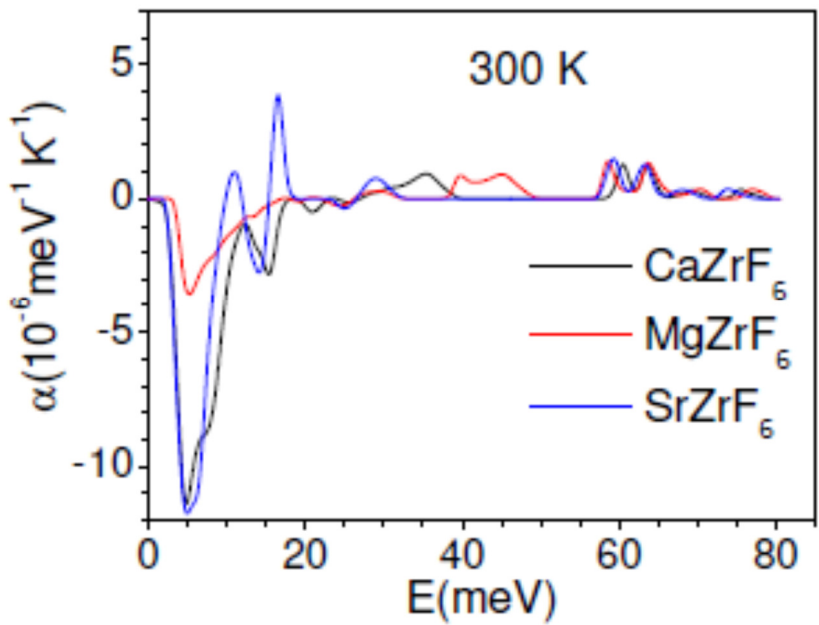

FIG. 6 (Color online) The calculated (solid line) and experimental(symbols)[21, 28] fractional change in volume with respect to the volume at 300 K in MZrF$_6$ (M=Ca, Mg and Sr) as a function temperature.

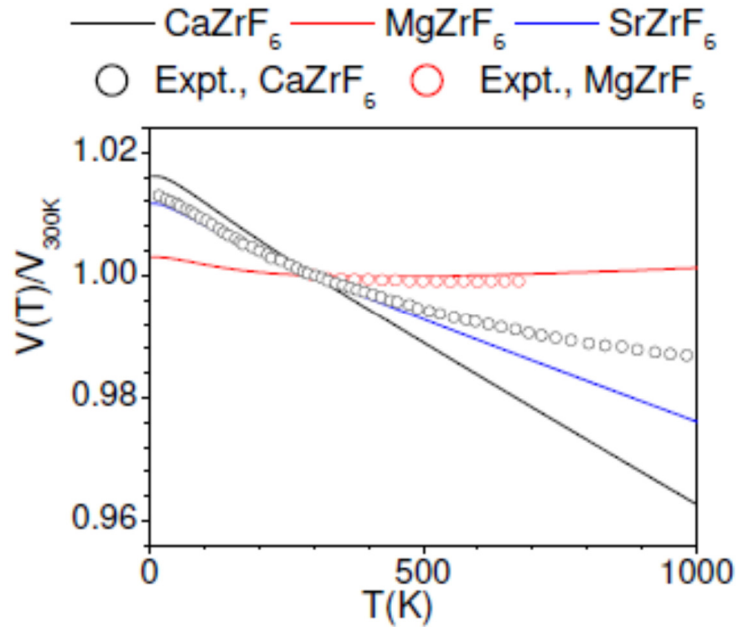

FIG. 7 (Color online) (a)The calculated mean square displacement of various atoms in MZrF$_6$(M=Ca, Mg and Sr) as a function of temperature. (b)The calculated anisotropic mean squared displacement of fluorine atoms, along ($u_\parallel^2$) and perpendicular ($u_\perp^2$) to the F-Zr-F bond, in MZrF$_6$ (M=Ca, Mg and Sr) compounds as a function of temperature.

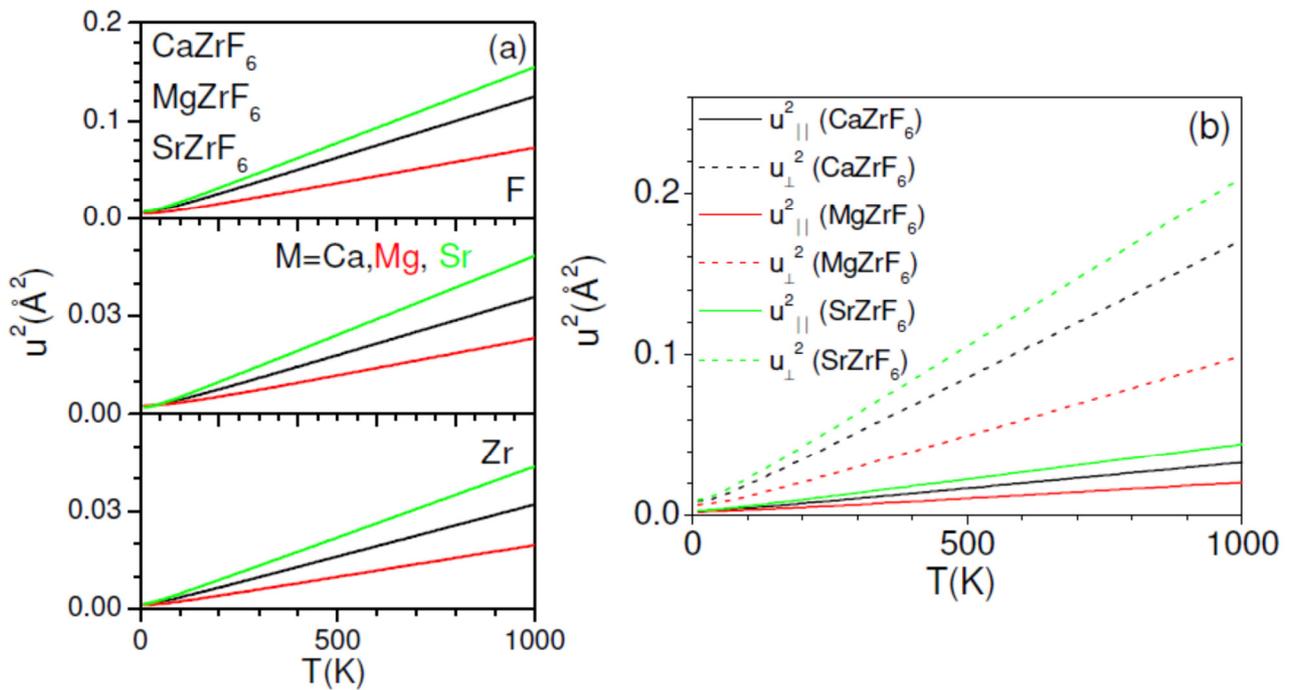

FIG. 8 (Color online) The displacement pattern of phonon modes having large negative Grüneisen parameter at Γ(0 0 0) and X(1 0 0) points in the Brillouin zone. The energy of the Γ(0 0 0) point mode is 3.6, 3.87 and 3.31 meV and that of the X point mode is 3.66, 3.95 and 3.51 meV for $CaZrF_6$, $MgZrF_6$ and $SrZrF_6$ respectively. The orange and green octahedral units correspond to $MF_6$ and $ZrF_6$ respectively.

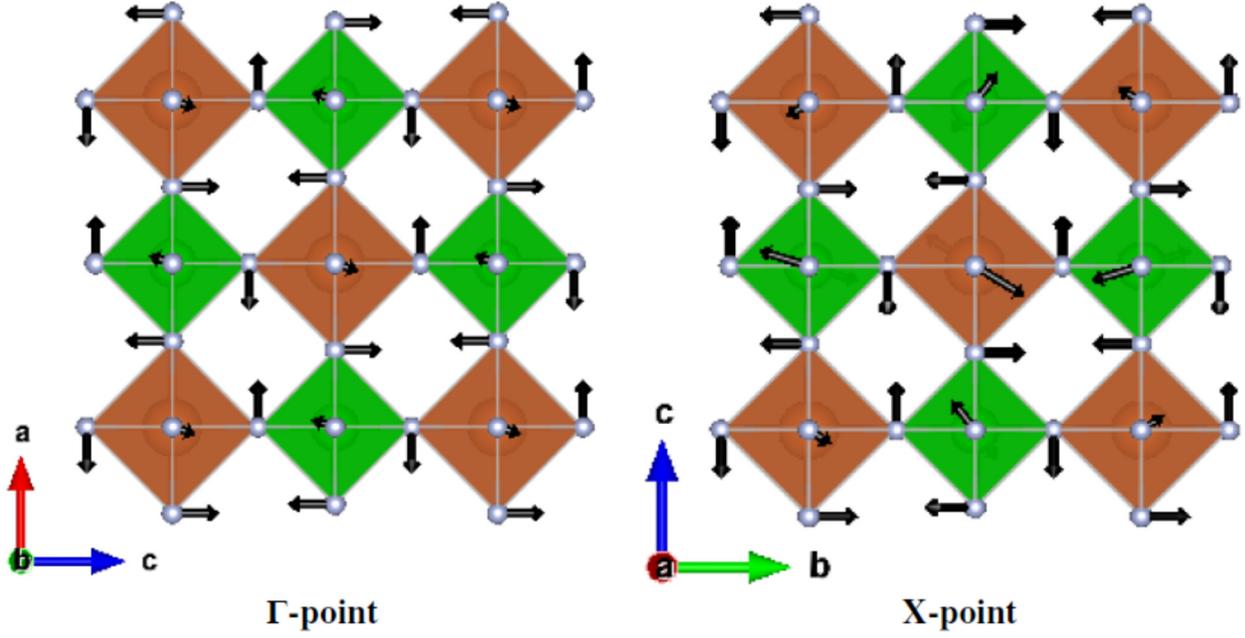

FIG. 9 (Color online) The calculated potential energy profile (black lines) of specific phonon modes at Γ(0 0 0) and X(1 0 0) points in the Brillouin zone. The red line shows the harmonic profile of respective phonon modes.

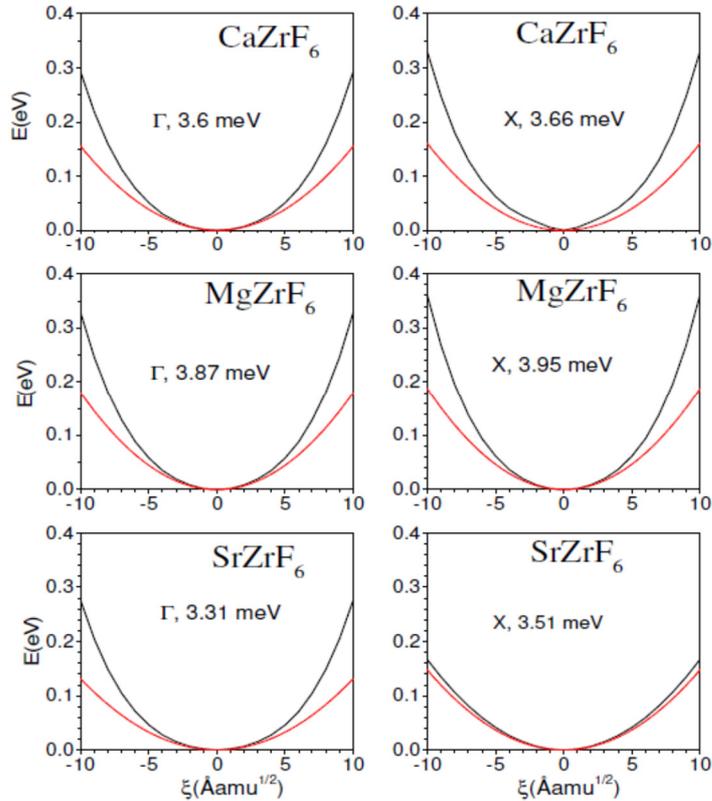

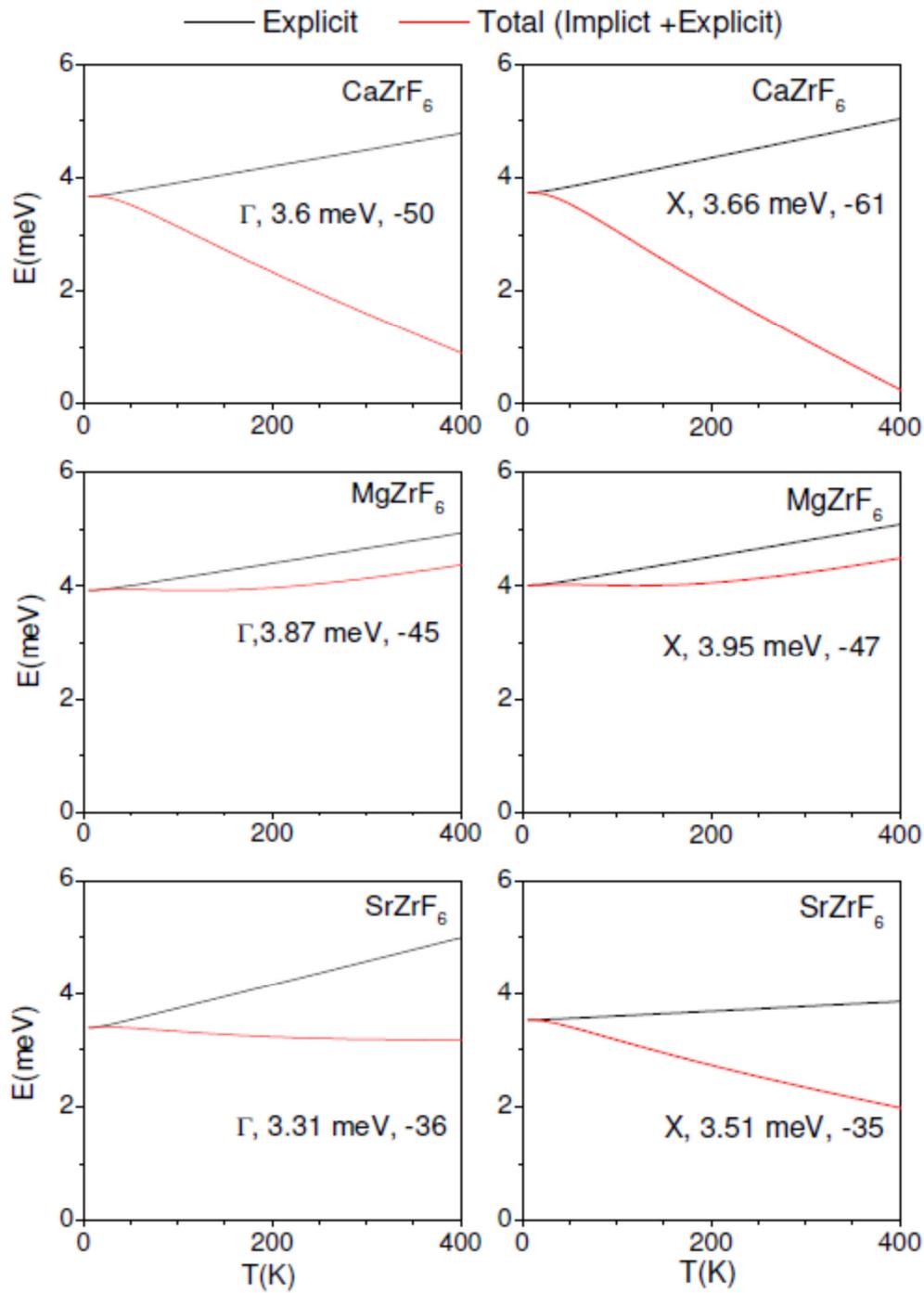

FIG. 10 (Color online) The calculated temperature dependence of specific phonon modes at Γ(0 0 0) and X(1 0 0) points in the Brillouin zone. The numbers next to the symmetry point are energy and Grüneisen parameter (Γ) of the phonon mode.